# The mitigating role of regulation on the concentric patterns of broadband diffusion. The case of Finland

Jaume Benseny[1], Juuso Töyli[1,2], Heikki Hämmäinen[1], Andrés Arcia-Moret[3]

[1]Communication and Networking Department, Aalto University,
Konemiehentie 2, P.O.Box 15400, Espoo, Finland

[2]Department of Economics, University of Turku,
Rehtorinpellonkatu 3, 20500, Turku, Finland

[3]Computer Laboratory, University of Cambridge,
William Gates Building, 15 J J Thomson Ave, Cambridge, UK

{jaume.benseny, juuso.toyli, heikki.hammainen}@aalto.fi, andres.arcia@cl.cam.ac.uk

**Abstract**

This article analyzes the role of Finnish regulation in achieving the broadband penetration goals defined by the National Regulatory Authority. It is well known that in the absence of regulatory mitigation the population density has a positive effect on broadband diffusion. Hence, we measure the effect of the population density on the determinants of broadband diffusion throughout the postal codes of Finland via Geographically Weighted Regression. We suggest that the main determinants of broadband diffusion and the population density follow a spatial pattern that is either concentric with a weak/medium/strong strength or non-concentric convex/concave. Based on 10 patterns, we argue that the Finnish spectrum policy encouraged Mobile Network Operators to satisfy ambitious Universal Service Obligations without the need for a Universal Service Fund. Spectrum auctions facilitated infrastructure-based competition via equitable spectrum allocation and coverage obligation delivery via low-fee licenses. However, state subsidies for fiber deployment did not attract investment from nationwide operators due to mobile preference. These subsidies encouraged demand-driven investment, leading to the emergence of fiber consumer cooperatives. To explain this emergence, we show that when population density decreases, the level of mobile service quality decreases and community commitment increases. Hence, we recommend regulators implementing market-driven strategies for 5G to stimulate local investment. For example, by allocating the 3.5 GHz and higher bands partly through local light licensing.

## 1. Introduction

Empirical studies indicate that broadband services have a positive impact on economic growth, employment, and productivity (Arvin and Pradhan, 2014; Koutroumpis, 2009; Minges, 2015). Hence, National Regulatory Authorities (NRAs) establish broadband penetration goals in their National Broadband Plans. For example, the United Kingdom aimed for 90% of households to have 2 Mbps download speed services by 2015. In comparison, Finland aimed for 100 Mbps download speed services to be made available no more than 2 kilometers away from permanent residences, according to demand. NRAs typically pursue penetration goals via multiple policies, including Universal Service Obligations (USOs) and spectrum auctions with coverage obligations. Although NRAs can impose USOs forcing operators to provide minimum services, they typically subsidize non-profitable investment through Universal Service Funds (USFs). USOs can distort competition since cost calculations do not account for intangibles, such as brand recognition and network effects (Alleman et al., 2010; Batura, 2016; Calvo, 2012). In addition, NRAs can allocate spectrum while demanding coverage obligations, balancing government revenues with coverage obligation investments (Cave and Hatta, 2008). However, spectrum auctions can also distort competition by establishing dominant spectrum holders or by raising entry barriers.

The above-mentioned policies not always succeed, often leaving rural and remote areas without access to modern broadband services (Prieger, 2013; Townsend et al., 2013). To address this service inequality, underserved communities can deploy broadband infrastructure and provide local coverage. During the 2000s while new services like VDSL and cable were deployed in European cities, DSL over long copper lines remained as the only solution for rural users. To increase the quality of rural broadband, wireless community networks emerged in Spain, Germany, Greece, among other European countries, thus showing the willingness of end users to invest in broadband infrastructure (Baig et al., 2015; Berger and Frey, 2016; Frangoudis et al., 2011; Fuchs, 2017; Micholia et al., 2018; Saldana et al., 2017). Therefore, broadband penetration goals not only can be achieved through nationwide operator investment but also considering the willingness and investment capacity of end users.

This article analyzes the role of Finnish regulation in achieving the broadband penetration goals defined by the NRA. Finland is an interesting study case since it defined ambitious penetration goals, pioneered in technology-neutral USOs, and pro-actively encouraged investment in mobile and fixed services. In comparison to EU countries, mobile broadband has experienced a rapid growth with consistently higher penetration rates and average traffic volumes (OECD, 2017). However, fiber deployments have been comparatively delayed (EC, 2015), provoking the emergence of fiber community networks.

It is well known that population density has a positive effect on broadband diffusion since higher densities typically imply favorable conditions for operator investment (e.g., economies of scale in service provisioning) and user adoption (e.g., higher education and income levels). We study the effect of population density on the determinants of broadband diffusion via Geographically Weighted Regression[1] (GWR) at the postal code level during 2013-2016 by combining socio-economic data from the Official Statistics of Finland and mobile network measurements from the Netradar.org. The 2013-2016 is an interesting period since it covers the time between the 4G auction of 800 MHz (in 2013) and the 95% coverage obligations deadline (in 2016). We suggest that the main determinants of broadband diffusion and the population density follow a spatial pattern that is either concentric with a weak/medium/strong strength or non-concentric convex/concave. Based on the 10 obtained patterns, we analyze the role of Finnish regulation. For example, we argue that if the concentric pattern on an individual diffusion determinant is strong, policies addressing this particular determinant have not been able to mitigate the effect of population density.

The rest of this article is structured as follows. Section 2 reviews the Finnish regulation on broadband, including universal service goals. Section 3 conducts a literature review on the effect of population density on broadband diffusion. Section 4 proposes the concentric patterns of broadband diffusion. While Section 5 describes the Finnish data under study, Section 6 presents the GWR method. Section 7 uncovers the effect of population density on diffusion determinants. Section 8 discusses the role of Finnish regulation expanding service coverage into rural areas. In Section 9, conclusions are drawn. Finally, Section 10 describes the study limitations and future research.

---

[1] GWR has been recently used to model spatial relationships that suffer spatial non-stationarity (Kamarianakis et al., 2008; Mitchell and Yuan, 2010).

## 2. Broadband regulation in Finland

Over the years, the Finnish NRA has consistently pursued a market-driven regulatory strategy with a very limited budget for subsidies. During the early DSL-dominated broadband market, no operator had significant market power since subscribers remained evenly distributed among regional operators. Moreover, potential dominant behavior was avoided by issuing cost-orientation requirements and price caps for DSL in 2005 (Eskelinen et al., 2008; Pursiainen, 2007). In 2008, Finland defined a new National Broadband Plan aiming to make 100 Mbps download speed services available no more than 2 kilometers away from permanent residences, according to demand. The NRA did not define how these last 2 kilometers should be covered or what minimum service quality should be finally provided to households. To this end, a combination of policies has been introduced encouraging investment in fixed and mobile services (Harri Pursiainen, 2008).

First, technology-neutral USOs were established in 2010 enforcing the provision of 1 Mbps download speed services to requesting permanent residences. Note that no USF was created and universal service providers were not compensated for connecting customers. USO's download speed was increased to 2 Mbps in 2015.

Second, to stimulate rural investment in new fixed services, a national project called "Broadband for all 2015" was launched to subsidize projects deploying fiber access in rural areas. However, 5 years later, the government had to relax the subsidizing conditions since the number of received projects did not cover for the deployment targets (FICORA, 2013; LVM, 2013a). Subsidies for fiber deployment have been received by nationwide, regional, and community operators. Examples of fiber consumer cooperatives are Optowest, EmsaloNet, KairanKiutu, Utakuiti, Siikaverkko, Kymijoki Village, Rautavaaran tietoverkko-osuuskunta (FICORA, 2013; LVM, 2015).

Third, to encourage investment in 3G mobile services, in 2009 the NRA allocated an equal amount of 2.6 GHz spectrum to the four competing MNOs. The same year, the Finnish regulator was the first European NRA to auction 1800 MHz bands, thus prompting an early and ambitious deployment for 4G infrastructure. More importantly, an equal amount of spectrum was again assigned to competing MNOs while demanding coverage obligations for 99% of the population. In the following years, additional spectrum from lower frequency bands was auctioned, including the 800 MHz band in 2013 and the 700 MHz band in 2016, stimulating investment in rural areas. Again, 800 MHz spectrum was equally assigned to MNOs while demanding 95% and 99% population coverage within 3 and 5 years, respectively. Note that the level of demanded coverage for the 800 MHz could be achieved with 4G infrastructure using other bands (LVM, 2013b). Furthermore, in 2014 the NRA allowed a joint venture between 2 out of the 3 MNOs for them to comply with 4G coverage obligations in Eastern and Northern Finland, which are the least populated areas in the country. This joint venture is allowed to operate radio infrastructure combining both MNOs' 4G spectrum bands. However, the venture is also required to rent our masts and sites to competitors (BEREC, 2018).

As a result of the Finnish regulation, high broadband penetration has been achieved fuelled by mobile which reached a 138% penetration rate in 2014. However, in 2015, rural fiber coverage in Finland remained below the EU average (EC, 2015).

## 3. Literature review on the effect of population density on broadband diffusion

The diffusion of broadband services has been extensively studied through contagious models which assume that the spreading phenomena occur through direct citizen contact. For example, the Bass diffusion model describes a product adoption process for a population that is divided between innovators and imitators (Bass, 1969). Similarly, the theory for the diffusion of innovations suggests that the spread of innovations depends on the innovation itself, the communication channels, time, and the social system (Rogers, 1971). More recently, both the adoption and diffusion of broadband services have been modeled through logistic and linear models including a large number of supply and demand determinants. Along the years, a short list of key determinants for the diffusion of broadband services has consolidated (Bauer et al., 2014; Glass and Stefanova, 2010; Lyons, 2014). In this section, we search for evidence on population density effect (either a positive or a negative effect) on diffusion determinant by reviewing literature that is specific to each determinant (e.g., PISA reports for education, techno-economic studies for access network costs per user).

Population density effect on operator investment

Operators assess the economic viability of infrastructure investments by estimating the cost per user of access networks. This cost is known to decrease as population density increases due to the economy of scale benefits in capital and operational expenses. The negative effect of population density on access network costs is identified for DSL in Ireland (Lyons, 2014) and fiber in Australia (Lee et al., 2009). The same effect is identified in techno-economic studies for WiMAX (Smura et al., 2007) and mobile networks (Johansson et al., 2004; Oughton and Frias, 2017). While fixed networks require dedicated ducts to bridge the last mile, wireless networks share spectrum among subscribers. Hence, access network costs per user are typically lower for wireless compared to fixed networks.

Wireless operators assess investment viability by forecasting spectrum usage since the available spectrum limits the maximum access capacity of base stations. Since base stations share a limited amount of spectrum among subscribers, the usage of spectrum typically increases with subscriber density (Jäntti et al., 2011). Therefore, wireless networks usually present a higher spectrum usage (in bits/s/MHz) as population density increases. This relationship is accepted in the literature for licensed (Clarke, 2014), and unlicensed spectrum (De Filippi and Tréguer, 2015; Kajita et al., 2014) .

The willingness of operators to invest also depends on the level of service competition. The motivation of operators to compete for users decreases as population density decreases, given that revenues per user may not compensate for the increase in access cost per user. In the case of mobile services, revenues per user can hardly increase as population density decreases since data plans are priced nationwide. In the case of fixed services, revenues per user might slightly increase with population density decrease, given that prices in rural areas are typically higher than in the urban ones. The motivation of operators to compete becomes even more sensitive to population density decrease when services are offered via flat-rate pricing, given that revenues per user are constant. The positive effect of population density on competition has been observed in OECD countries, where competition levels were found to remain higher in cities compared in rural areas (Berger and Frey, 2016). This positive effect has also been observed in the US, where the number of Internet Service Providers (ISPs) in the US counties was found to statistically decrease as population density decreases in 2007 (Durairajan and Barford, 2016; Grubesic, 2010).

Operator investment can also be encouraged through state subsidies, thus directly alleviating the operator costs in low-profit areas. For example, EU regulation prioritizes the allocation of state aid by classifying areas as white (no providers), grey (1 provider), or black (2+ providers) (OECD, 2011). Based on this example, it appears that state subsidy is influenced by population density like service competition but with an opposite sign. Since service competition is positively influenced by population density, the state subsidy is likely negatively influenced in line with studies arguing for the allocation of public funds to the deployment of next-generation fixed access in markets with geographical constraints (Ragoobar et al., 2011).

As demonstrated by the emergence of European wireless community networks, local actors may invest in broadband infrastructure when local demand remains underserved and feels aggravated. However, the viability of local broadband projects is highly dependent on the level of community commitment (Frangoudis et al., 2011; Micholia et al., 2018). In any case, the commitment of communities to self-governance is typically higher in rural compared to urban areas. For example, rural small businesses are usually managed according to rural socio-cultural values in which the word of mouth reputation, as well as primacy of relations with family, friends, and neighbors, play a significant role (Shields, 2005). Similarly, higher levels of social cohesion are typically found in rural compared to urban areas, given the higher percentage of households living in owner-occupied dwellings and lower levels of immigration.

Population density effect on user adoption

The willingness of users to adopt broadband services is typically associated with their level of education. In this context, learning opportunities are typically associated with population density and the labor market. For example, students in city schools of OECD countries outperform those in rural areas by 40 score points on average, or the equivalent of one academic year (OECD, 2010). Furthermore, prospects for better jobs typically drive the rural population to urban areas. Therefore, we suggest that education is positively influenced by population density.

In addition to education, household income has also been associated with user adoption, given that both fixed and mobile services are typically acquired at the household level. According to (Eurostat, 2014), in 2012, the fraction of EU population at risk of poverty was 27.3% in thinly-populated (rural) areas, 22.6% in intermediate density areas, and 24.7% in densely-populated (urban) areas. This information suggests that the effect of population density on household income is not monotonous but follows a concave shape. Indeed, urban empirical studies show that the population is not only distributed according to socioeconomic characteristics, but also to family life-cycle, and ethnic divisions (Murdie, 1969). As a result, spatial patterns in the form of high-low income neighborhoods might appear.

User adoption can also be stimulated through private subsidies, including MNO subsidy, mobile advertisement, and free Wi-Fi. Urban-to-rural subsidization by MNOs increases with the decrease of population density. Given that prices for mobile data plans are the same nationwide, revenues per mobile user remain roughly constant regardless of the user location. Since access cost per user increases as population density decreases, MNOs need to subsidize rural users with revenues from the urban ones to satisfy coverage obligations. Concerning mobile advertisement, private companies implementing advertisement-based revenue models might grant mobile data to users through multiple approaches. In the earned mobile data model, data is granted to specific user profiles in exchange for actions through third-party applications (e.g., Gigato, mCent). However, user profiling is not necessarily associated with population density. In data plans with zero-rated applications, data is granted to subscribers regardless of location (Dhanaraj Thakur, 2016). In addition to mobile advertisement, free Wi-Fi access can also be considered a private user subsidy. However, there is no reported empirical evidence on the relation between free Wi-Fi access and population density.

As described by contagious diffusion models, the more exposed and knowledgeable a user becomes with a service, the more likely the adoption. Hence, we suggest that service penetration is positively influenced by population density since broadband services typically present higher penetration rates in urban areas compared to rural areas. For example, DSL coverage in the EU27 countries by the end of 2008 was only available to 77% of the rural population compared to the 92% national average. As a result, urban and rural penetration rates reached 18% and 12.3%, respectively (EC, 2010). Similarly, the availability of fixed services, including DSL and cable, was found to be lower for rural locations in the US (Prieger, 2003). This difference in penetration rates is also observed for mobile services. By the end of 2008, in the EU27 countries, 3G coverage was available to 74% of the total population and 38% of the country area. Although 3G penetration reached 12.4% of the total population, users mainly lived in cities and dense suburban areas.

The willingness of users to adopt broadband services is also associated with service quality. Some regulators organize measurement campaigns to validate the correspondence between delivered and advertised quality of service (QoS). Although these campaigns may cover areas with varied population density, the published statistics are limited to weighted nationwide averages (OECD, 2014). Nevertheless, some regulatory agencies individually report annual comparative values. For example, average fixed service speeds in the UK were more than three times faster in urban than in rural areas in 2016 (OFCOM, 2017). Also in the UK, mobile broadband service presents QoS disparities between rural and urban areas (Speedtest, 2018; Townsend et al., 2013).

Table 1 summarizes the above-cited evidence about the one-to-one effect of population density on diffusion determinants, which has been collected by reviewing literature that is specific to each determinant. No cross-effect between determinants is implied.

*Table 1. Population density effect*

| Diffusion determinants | Fixed services | Mobile services |
|---|---|---|
| Access network cost per user | **negative**<br>DSL (Lyons, 2014)<br>Fiber (Lee et al., 2009)<br>WiMAX (Smura et al., 2007) | **negative**<br>Mobile (Johansson et al., 2004) |
| Spectrum usage | **positive**<br>Unlicensed spectrum<br>(De Filippi and Tréguer, 2015)<br>(Kajita et al., 2014) | **positive**<br>4G bands (Clarke, 2014) |
| Competition | **positive**<br>Number of operators<br>(Grubesic, 2010)<br>(Durairajan and Barford, 2016) | **positive**<br>(no source) |
| State subsidy | **negative**<br>EU state aid regulation (OECD, 2011) | |
| Community commitment | **negative**<br>Social cohesion (Shields, 2005)<br>Municipality support (Baig et al., 2015). | |
| Education | **positive**<br>Secondary school (OECD, 2010) | |
| Income | **concave**<br>Household income (Eurostat, 2014) | |
| Private user subsidy | **unknown**<br>Free Wi-Fi access<br>(no source) | **negative**<br>MNO subsidy (Peltola and Hämmäinen, 2018)<br>**non-concentric**<br>Ad-based (Dhanaraj Thakur, 2016) |
| Service penetration | **positive**<br>Rates for DSL (EC, 2010) | **positive**<br>Rates for 3G (EC, 2010) |
| Service quality | **positive**<br>Download speed (OFCOM, 2017) | **positive**<br>Download speed (Speedtest, 2018) |

## 4. Modeling broadband diffusion through concentric patterns

In urbanization studies, it is generally accepted that population density decreases with radial distance from the city center. According to recent studies, this decrease is best modeled through an inverse power function since it embodies features from fractal geometry, improving the fit for urban form and density (Batty and Sik, 1992; Smeed, 1963). Therefore, we assume that population density decreases with radial distance from city centers following a negative power model, as shown in Figure 1A. As a result, we observe that the population is distributed in concentric circles of decreasing population density, as shown in Figure 1B. Based on these assumptions, we argue about the effect of population density on the diffusion of broadband services by proposing a concentric pattern for each diffusion determinant. To be precise, we identify a diffusion determinant following a concentric pattern when this determinant and the population density share a central point from which the determinant consistently increases or decreases with decreasing population density (i.e., the relationship between the population density and the determinant is monotonic).

We define the concentric pattern to be positive when the determinant and population density increase or decrease simultaneously (i.e., the relationship between the two follows a monotonically increasing function) and to be negative in the alternative case (i.e., the relationship between the two follows a monotonically decreasing function). In any other circumstances, we consider the effect to be non-concentric either convex or concave. Figure 1C shows an example of positive concentric pattern in green color and an example of a negative concentric pattern in red color. Finally, we adapt the findings of the literature review on the effect of population density to comply with our definition of a concentric pattern, as presented in Table 2.

*Figure 1. Definition of the concentric pattern*

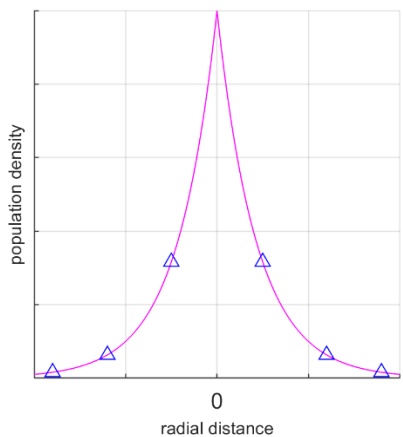
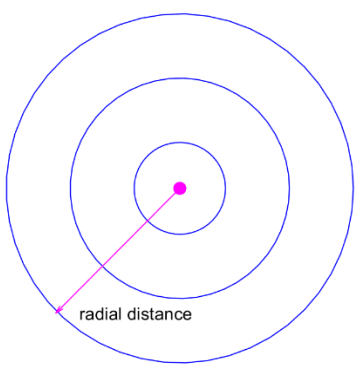
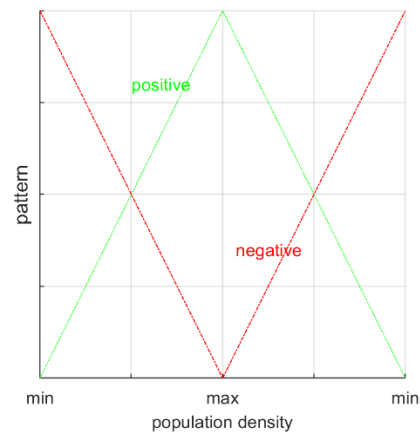

*Figure 1A.*
*Population density decrease with radial distance*

*Figure 1B.*
*Concentric circles of decreasing population density*

*Figure 1C.*
*Positive and negative concentric patterns (conceptual description)*

*Table 2. Concentric patterns of broadband diffusion*

| Diffusion determinants | **Concentric pattern sign** |
|---|---|
| Access network cost per user | - |
| Spectrum usage | + |
| Competition | + |
| State subsidy | - |
| Service quality | + |
| Education | + |
| Income | Non-concentric and concave |
| Private user subsidy | + / - / Non-concentric |
| Community commitment | - |
| Service penetration | + |

## 5. Data

Diffusion determinants are studied in Finland via proxy variables extracted from two data sets covering population socio-economics and network measurements.

Population socio-economics is extracted from the open Paavo repository produced by Statistics Finland (Official Statistics of Finland, 2017). This repository presents statistics at the postal code level including the 3036 postal codes of mainland Finland. From this repository, we extract the median household income (*med_hh_income*), the percentage of the population with a bachelor's degree (*%_bachelors*), and the percentage of households living in owner-occupied dwellings (*%_dw_ownership*), as proxies for income, education, and community commitment. We select median values of household income because this statistic is resistant to skewed distributions. We select bachelor's degrees (3 to 4 years after upper secondary education), given the equal learning opportunities in the Finnish education system. Further, we select the percentage of households living in owner-occupied dwellings as a proxy for community commitment since this variable has traditionally been associated with social inclusion, which is a necessary condition for community commitment. Finally, from this same repository, we also extract population density values for each postal code as well as polygon features supporting the computation of spatial statistics. Table 4 presents statistics, including cardinal numbers, for proxy variables and for the area of postal codes. Limitations caused by the lack of data and by the changing size of postal codes are discussed in Section 10. Summarizing, the average population density of Finland is 16 inhabitants per square kilometer, and approximately one-fourth of the total 5.5 million population is concentrated in the Helsinki metropolitan area. In the next page, Figure 2 depicts the postal codes of Finland using a choropleth map with a blended hue color progression, maximizing information visualization. Postal codes are grouped following a geometrical interval classification, thus balancing the number of postal codes per class versus interval length. To show the population density of small-sized postal codes, Figure 3 depicts the main centers of population density in more detail. These centers of population density will be referenced later in the results section.

Mobile network measurements are extracted from the Nettitutka.fi[2] repository (Nettitutka.fi, 2016). These are crowdsourced measurements obtained by the Nettitutka and Netradar mobile applications, which are available in the Android Play Store. The measurement service was first launched in Finland in 2012 and globally in 2013, achieving more than 7.8E6 measurements and 2.6E5 installations as of 10/2015 the majority of which occurred in Finland (Sonntag et al., 2013). To ensure that measurements were conducted via telecom infrastructure specifically deployed to serve residential population instead of roads, we discard measurements from devices moving faster than 3 m/s. For each postal code, we extract the number of observed mobile broadband providers (*MBP_number*) as proxy for competition, the median download speed for mobile (*med_speed_mobile*) as proxy for service quality, and the difference in days between the date when 4G services was first measured in Finland and in the postal code (*4G_diffusion_delay*) as proxy for service penetration. While measurements for *MBP_number,* and *med_speed_mobile* belong to 2015, the ones for *4G_diffusion_delay* cover the 3-year period between 04/2013 and 04/2016. To ensure that *4G_diffusion_delay* does not include erroneous measurements (e.g., from a malfunctioning device), we consider the date when 4G was first measured as a valid date, only when a second 4G measurement exists within a week time. When counting the number of mobile broadband providers per postal code, *MBP_number* accounts for Finnish operators, including both MNOs and mobile virtual network operators (MVNOs). In the case of *med_speed_mobile*, sample overrepresentation from individual users are alleviated via random sampling per unique installation. Table 4 presents statistics, including cardinal numbers, for proxy variables. Limitations caused by the lack of data and by sample selection effects are discussed in Section 10. Spatial matching of measurements to postal codes is performed using the GeoTrellis geospatial open-source library for Scala.

---

[2] Nettitutka.fi is the non-commercial version of the Netradar.org service.

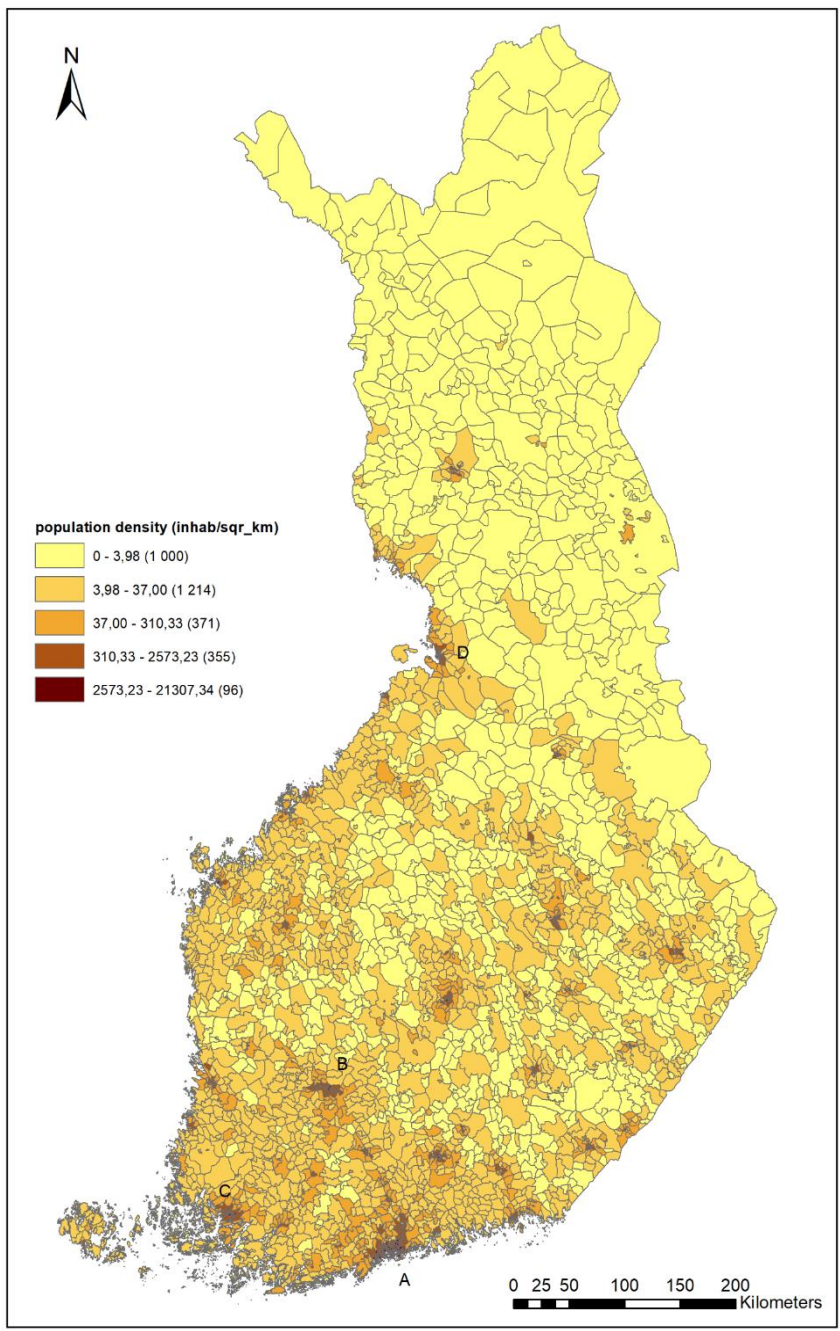

*Figure 2. Population density in the postal codes of Finland*

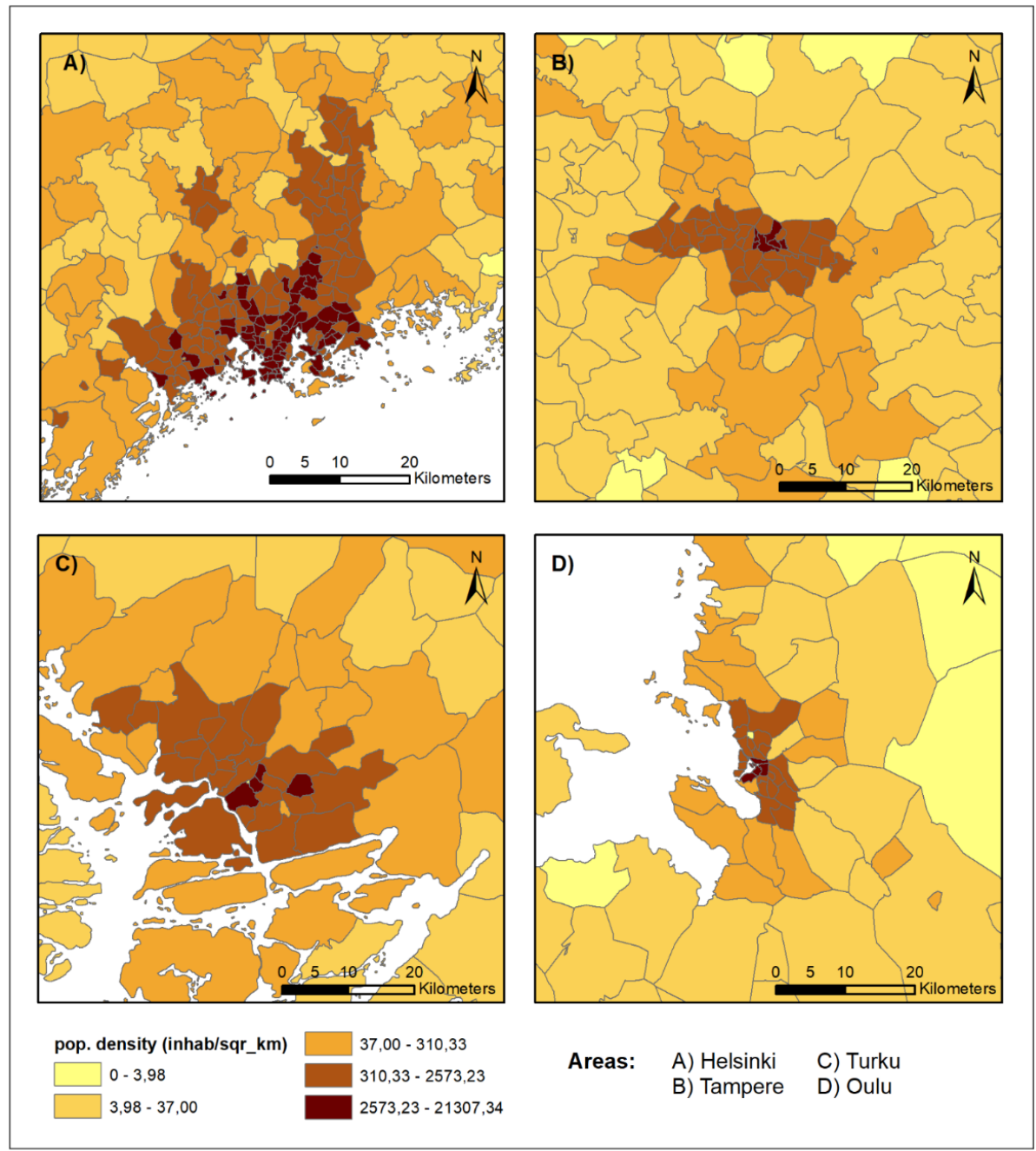

*Figure 3. Main centers of population density in Finland*

Table 3 summarizes the selected proxies for diffusion determinants and time spans.

*Table 3. Proxy variables for the study of Finland*

| Variable/Determinant | Proxy code | Time Span |
|---|---|---|
| Population density | *pop_dens* | 2014 |
| Competition | *MBP_number* | 2015 |
| Community commitment | *%_dwelling_ownership* | 2012 and 2017 |
| Education | *%_bachelors* | 2013 |
| Income | *med_hh_income* | 2013 |
| Penetration | *4G_diffusion_delay* | 04/2013 – 04/2016 |
| Service quality | *med_speed_mobile* | 2015 |

Table 4 reports basic statistics and measures of normality for the proxy and additional variables. We present values of skewness indicating that variable distributions are: skewed left for negative values, symmetric for zero values, and skewed right for positive values. In addition to skewness, we test distributions for the presence of heavy tails via the kurtosis test. While values of kurtosis tests below three indicate the presence of a light-tailed distribution, values above three indicate the presence of heavy tails.

*Table 4. Statistics for proxy and additional variables*

| Proxy code | Units | Min | Average | Max | Std. Dev. | Cardinal | Skewness | Kurtosis |
|---|---|---|---|---|---|---|---|---|
| *pop_dens* | inhab/km2 | 0 | 288.9 | 21,307.3 | 1,017.7 | 3,036 | 8.0 | 107.8 |
| *MBP_number* | operators | 1 | 2.9 | 6 | 1 | 2,849 | -0.4 | 2.1 |
| *%_dw_ownership* | percentage | 0 | 0.814 | 1 | 0.16 | 2,848 | -1.5 | 3.9 |
| *%_bachelors* | percentage | 0 | 5 % | 75 % | 4 % | 2,925 | 3.7 | 33.7 |
| *med_hh_income* | euros / hh | 0 | 33,480 | 72,889.0 | 8,069.3 | 2,868 | -0.9 | 5.6 |
| *4G_diffusion_delay* | days | 0 | 548.6 | 1,090.0 | 263.3 | 2,201 | -0.2 | 2.2 |
| *med_speed_mobile* | kbps | 15.0 | 8,696 | 67,551.0 | 6,320.6 | 1,924 | 1.9 | 10.3 |
| Additional variables | | | | | | | | |
| *postal code area* | km2 | 0.614 | 111.2 | 8,378 | 282.4 | 3,036 | 13.98 | 310.4 |

## 6. Geographically weighted regression (GWR)

Since the data under study covers an entire country, the relationship between population density and the diffusion determinants may present spatial heterogeneity. For example, postal codes located in northern Finland might present lower levels of household income than postal codes in southern Finland, albeit they have similar levels of population density. To address the spatial non-stationarity of global coefficient estimates, the geographically weighted regression (GWR) method has been proposed to measure local relationships between variables that differ from location to location (Fotheringham et al., 1998). GWR extends the method of ordinary least squares (OLS) by allowing coefficients to vary locally, becoming location-specific rather than global estimates. Suppose a series of independent variables $\{x_{ij}\}$ and a dependent variable $\{y_i\}$ where $i$ = 1,2,…m, and $j$ = 1,2,….n. In OLS, global coefficients are estimated according to:

$$y_i = C_0 + \sum_{j=1}^{n} C_j x_{ij} + \varepsilon_i$$

where $y_i$ is the value of the dependent variable $y$ at location $i$, $C_0$ is the intercept, $C_j$ is the coefficient for the independent variable $x_{ij}$. In addition, $\varepsilon_i$ represents the error term, which is generally assumed to be independent and normally distributed with 0 means and constant variance $\sigma^2$. This type of regression is considered global since coefficient estimates are assumed to present spatial stationarity. In contrast, the GWR method recognizes that the coefficient estimates in a regression model can vary throughout the space, thus including spatial non-stationarity in the relationship between the dependent and independent variables. GWR estimates local coefficients according to:

$$y_i = C_0(u_i, v_i) + \sum_{j=1}^{n} C_j(u_i, v_i) x_{ij} + \varepsilon_i$$

where $(u_i, v_i)$ is the spatial location of the $i$-th observation and $C_j(u_i, v_i)$ is the value of the $j$-th parameter at location $i$. The coefficients of this equation are estimated at each location $i$. The GWR coefficients are typically estimated using the weighted least squares method, which weights the values of neighboring locations through a distance decay function. A key parameter in GWR is the bandwidth parameter. The bandwidth indicates the radial distance from which the weighting function becomes 0, thus limiting the number of locations included in the coefficient estimate for the location $i$. In this study, we measure the effect of population density on each diffusion determinant by estimating the local coefficients $C_0(pc_i)$ and $C_1(pc_i)$ according to:

$$determinant_i = C_0(pc_i) + C_1(pc_i)\, pop\_dens_i + \varepsilon_i$$

where $pc_i$ is the postal code of the $i$-th observation. To this end, weighted least squares is conducted for each postal code and the value of nearby postal codes according to a Gaussian curve. When estimating GWR coefficients for postal code $i$, the value at postal code $j$ is weighted through $w_{ij}$ which is calculated as:

$$w_{ij} = \exp\left(\frac{d_{ij}}{\beta}\right)^2$$

where $d_{ij}$ is the distance between postal code $i$ and postal code $j$ and β is the bandwidth parameter. Given the variance in the spatial density of Finnish postal codes, we decide to employ an adaptive bandwidth. In contrast to fixed bandwidths, which indicate a fixed distance, adaptive bandwidths indicate a number of neighboring areas. By employing an adaptive bandwidth, the weighted least squares calculation includes postal codes that are further away from postal code $i$ when this code is found in a low-density region, and it employs smaller distances when postal code $i$ is found in a high-density region. We obtain a determinant-specific optimal adaptive bandwidth though model cross-validation. Analysis were done in ArcMap Desktop 10.4 (ESRI, n.d.).

## 7. Results

### 7.1. GWR results

Table 5 presents GWR results in goodness-of-fit and spatial behavior. Results from the Moran´s I test show that regression residuals for *MBP_number* and *%_bacherlors* suffer spatial autocorrelation. The existence of spatial autocorrelation of residuals indicates that population density alone cannot explain the observed variance in the dependent variable. Hence, the regression model is miss-specified and misses an explanatory variable, invalidating coefficient estimates. In the following subsection, we elaborate on how these results are interpreted. Apart from these two regression models, values of $R^2$ range between 0.31 and 0.52, which we consider acceptable since population density is the only explanatory variable in our models. Further, we observe low bandwidth values that vary between 64 and 86 postal codes, which suggests that coefficient estimates might present high spatial variance but low local bias (Charlton and Fotheringham, 2009). In this context, Table 5 reports effective numbers enabling study reproducibility[3].

*Table 5. GWR goodness-of-fit and spatial behavior*

| Determinant | Units | Moran's I (p-value) | $R^2$ | Observations (postal codes) | Bandwidth (postal codes) | Effective Number |
|---|---|---|---|---|---|---|
| *MBP_number* | operators | 0.02** | 0.36 | 2,849 | 79 | 202.68 |
| *%_dw_onwership* | percentage | 0.13 | 0.50 | 2,848 | 81 | 198.40 |
| *%_bachelors* | percentage | 0.01*** | 0.68 | 2,925 | 78 | 212.12 |
| *med_hh_income* | euros/hh | 0.68 | 0.38 | 2,868 | 86 | 187.69 |
| *4G_diffusion_delay* | days | 0.11 | 0.52 | 2,201 | 58 | 214.43 |
| *med_speed_mobile* | kbps | 0.86 | 0.31 | 1,924 | 64 | 198.32 |

p-values that are statistically significant are indicated as follows: 1% ***, 5% **

Table 6 presents GWR results concerning coefficient estimates and standardized residuals. Coefficient estimates are described via the five-number summary, given their skewed distribution. Standardized residuals are described via extreme and average values since their distribution is typically centered in zero with some tail deviations. These are analyzed in detail in the next subsection.

*Table 6. GWR coefficient estimates and standard residuals*

| | Coefficient estimate $C_1$ | | | | | Standardized residuals (Std.Res.) | | |
|---|---|---|---|---|---|---|---|---|
| Determinant | Min. | p25 | p50 | p75 | Max. | Min. | Avg. | Max. |
| *MBP_number** | -0.12 | 4.19E-04 | 1.40E-03 | 8.04E-03 | 2.24 | -3.27 | -0.01 | 3.09 |
| *%_dw_onwership* | -0.039 | -5.10E-04 | -2.60E-04 | -1.60E-04 | 2.00E-6 | -6.84 | 0.00 | 3.72 |
| *%_bachelors** | 1.48E-4 | 1.37E-05 | 3.09E-05 | 5.01E-05 | 0.01 | -5.46 | -0.05 | 27.70 |
| *med_hh_income* | -153.62 | -6.11 | -4.15 | -2.11 | 2.49E03 | -8.35 | -0.03 | 5.06 |
| *4G_diffusion_delay* | -25.72 | -0.59 | -0.30 | -0.15 | 2.207 | -3.32 | 0.05 | 3.59 |
| *med_speed_mobile* | -68.14 | 1.75 | 4.66 | 8.94 | 316.52 | -3.92 | -0.04 | 6.99 |

p25/p50/p75 indicate the lower/median/upper percentiles, respectively.
* indicates that coefficient estimates cannot be trusted due to spatial autocorrelation of residuals

In addition to Table 6, the spatial variation of GWR coefficients is depicted across postal codes by Figures 4A, 4B, 4C, 4D. In these figures, postal codes are grouped using a geometrical interval classification, thus balancing the number of postal codes per class versus interval length. Given the different distributions among GWR coefficients, each figure uses a different classification which has been selected to maximize information visualization. The figures assign classes to color gradations by employing a common choropleth map with a single hue progression. The number of postal codes per class is indicated in parenthesis next to the class range. These figures are analyzed in detail in the next subsection.

---

[3] We do not report Akaike Information criterion (AICc) values since we do not compare goodness-of-fit between GWR models for the same determinant. Moreover, model complexity is limited by a single explanatory variable. We do not report tests for collinearity since our GWR models only include one single explanatory variable. Nevertheless, all GWR models present condition number values well below 30.

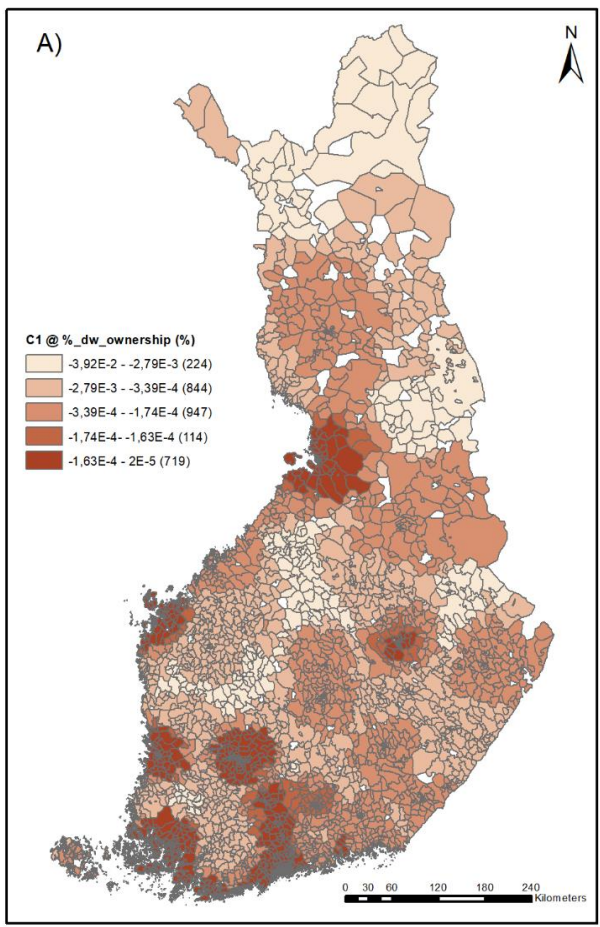
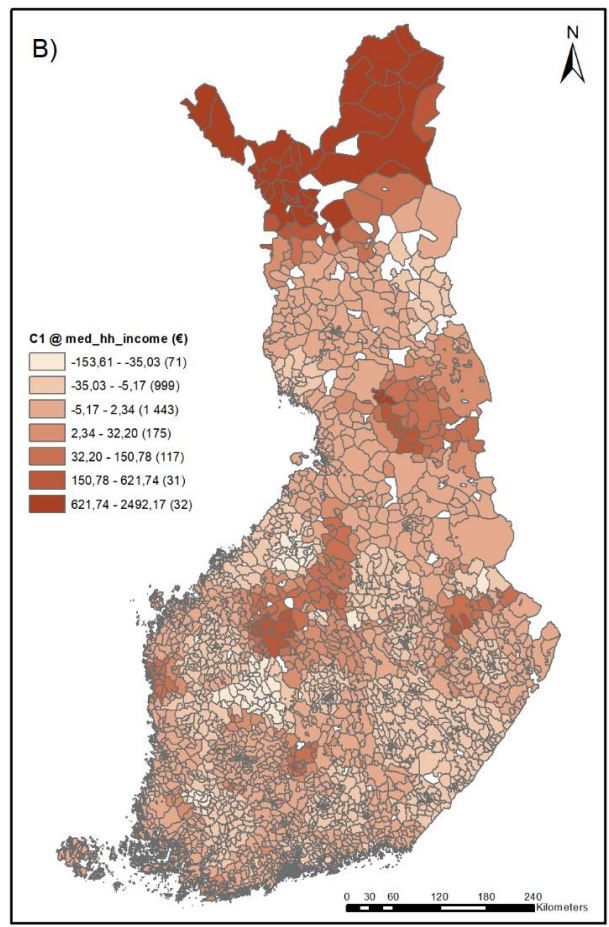
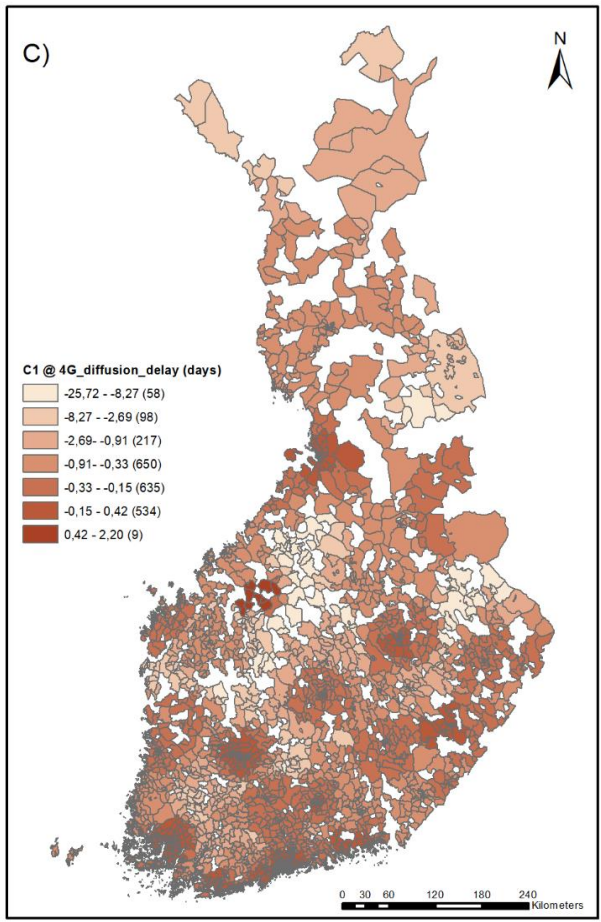
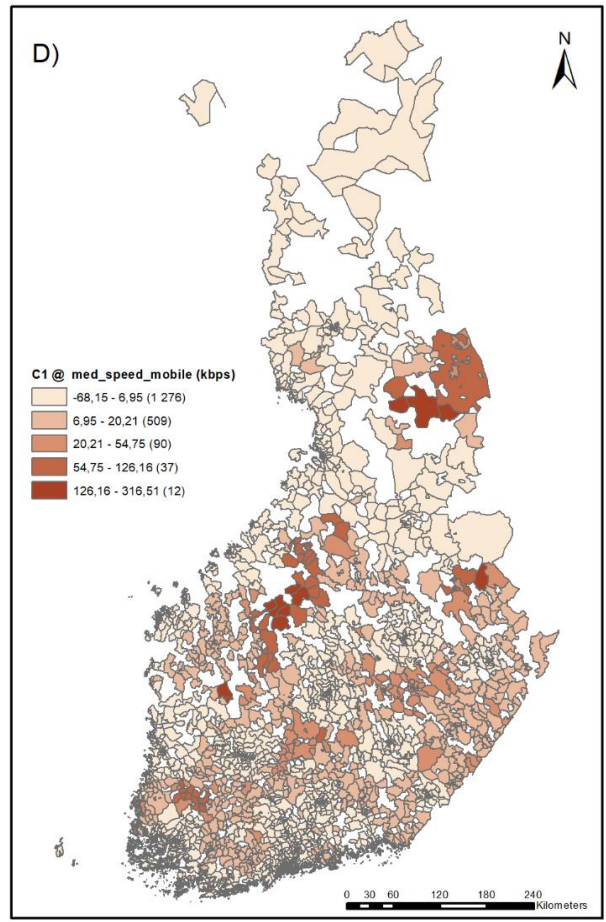

*Figure 4. Spatial variation of GWR coefficients*

## 7.2. Measuring the concentric patterns of broadband diffusion

This section describes how the GWR results are interpreted, allowing the authors to propose the existence of concentric patterns and their strength. We first assess the effect of population density by examining the spatial autocorrelation of GWR residuals through the Moran's I test. The existence of spatial autocorrelation of residuals indicates that population density alone cannot explain the observed variance in the dependent variable, i.e., the model is miss-specified and misses an explanatory variable. In this case, we consider the effect of the population density on the determinant to be weak. Second, if the GWR residuals do not suffer spatial autocorrelation, we assess the effect of population density to be concentric only if the GWR coefficients do not experience sign changes, indicating that the relationship between population density and the determinant is monotonic. We consider the concentric effect to be positive or negative according to the GWR coefficient sign. If the sign of GWR coefficient changes, we assess the effect as non-concentric either concave or convex. Third, we assess the effect strength of population density by examining the goodness-of-fit obtained by the GWR models through the coefficient of determination, a.k.a., $R^2$. Like OLS, GWR produces $R^2$ values with variance explanatory power, informing about the fraction of variance that can be explained by population density. We consider that the effect strength is strong when the model can explain at least 50% of the determinant variance, and the effect strength is medium when the explanatory power is less than 50%. The effect strength of the population density equals the strength of the concentric pattern.

After applying the interpretation principles described above, we suggest the existence of the following concentric patterns of broadband diffusion:

<u>Access network cost per user:</u> Since we lack operator cost data, we assume that in Finland access network costs per user decrease with increasing population density due to the economy of scale benefits, as identified for other countries in the literature review. Therefore, we suggest a negative concentric pattern.

<u>Spectrum usage</u>: Since we lack spectrum usage data, we refer to existing publications. Studies assessing the utilization of TV spectrum for mobile communications show that mobile spectrum usage in Finland increases as population density increases (Jäntti et al., 2011; Kerttula et al., 2012). Regarding the unlicensed spectrum, the study of the ISM band in the Oulu municipality revealed that spectrum usage was maximum in the city airport and city center. In contrast, usage was lower in two residential locations and non-existent in a peripheral rural area (Kokkoniemi and Lehtomäki, 2012). Therefore, we suggest a positive concentric pattern.

<u>Competition:</u> GWR results show that the number of mobile broadband providers per postal code (*MBP_number*) cannot be sufficiently explained by population density alone since the GWR residuals present spatial autocorrelation (Moran's I p-value: 0.02). Hence, we suggest that, in Finland, population density has a weak effect on mobile service competition. Therefore, we suggest a weak positive concentric pattern.

<u>State subsidies:</u> Although Finland adopted universal service obligations for broadband, these obligations are technology neutral, limited to 58 peripheral postal codes, and not supported by a USF. Therefore, universal service providers are not forced to use a specific broadband technology; neither are compensated for connecting costumers. Nevertheless, in parallel to USO, subsidies have been assigned to nationwide, regional, and community operators to build fiber access networks in areas where 100 Mbps download speed services are not available. Therefore, in principle, the availability of state subsidies should indicate the same population density effect than service quality but with an opposite sign. Since later in this section the population density effect on service quality is found to be non-concentric and concave, the effect should be convex for state subsidies. However, it is unlikely that state subsidies are allocated to profitable areas with high subscriber density. Hence, we conclude that the availability of state subsidies is influenced by population density in a negative concentric manner.

<u>Community commitment:</u> GWR results show that population density has a strong effect on the percentage of households living in owner-occupied dwellings (*%_dw_ownership*) since it can explain 50% of the observed variance ($R^2$: 0.5). Further, we consider this effect to be negative concentric since the coefficient estimate $C_1$ remains with a negative sign for the majority of postal codes, as shown in Table 6. In addition, Figure 4A shows how the strength of the population density effect decreases as postal codes become closer to the main centers of population density, where the *%_dw_ownership* reaches its minimum value. A closer look at Figure 4A reveals that $C_1$ values only become positive for a small subset of postal codes with the highest population density. Note that this small subset produces the absolute largest GWR standard residuals (Minimum Std. Res. < -3).

<u>Education:</u> GWR results show that the percentage of citizens with a bachelor's degree (*%_bachelors*) cannot be sufficiently explained by population density alone since the GWR residuals present spatial autocorrelation (Moran's I p-value: 0.01). Hence, we suggest that in Finland population density has a weak effect on education levels. Hence, we suggest a weak positive concentric pattern. This result is acceptable since the Finnish education system provides equal learning opportunities to students. For example, education is free of charge from elementary school to the university level.

<u>Income:</u> GWR results indicate that population density can explain 38% of the observed variance ($R^2$: 0.38) in median household income (*med_hh_income*). Nevertheless, this effect cannot be considered concentric since $C_1$ presents high variation, ranging from positive to negative values, as shown in Table 6. In more detail, Figure 4B depicts how the population density effect changes its sign from positive to negative as postal codes become closer to the main centers of population density. This result is coherent with findings from the literature review indicating that household income does not follow a monotonic function with the radial distance from the city center but typically peaks at a relatively short distance. Note that the number of postal codes with a negative $C_1$ is significantly larger than the opposite case. Regarding GWR standard residuals, the absolute maximum values are produced by postal codes with very high population density levels (Maximum Std. Res. > 3 and also Minimum Std. Res. < -3). Hence, we consider the population density effect to be non-concentric and concave.

<u>Private user subsidies</u>: Data plans with zero-rated applications are not available in Finland. Neither there are third-party mobile applications granting earned data either. Nevertheless, MNOs with coverage and uniform pricing obligations subsidize rural user costs with urban user revenues (Peltola and Hämmäinen, 2018). This subsidization is maximum for connections served via universal service obligations, given that universal service providers are not compensated. Hence, we suggest a negative concentric pattern.

<u>Service penetration:</u> GWR results show that population density has a strong effect on the delay of 4G diffusion (*4G_diffusion_delay*) since it can explain 52% of the observed variance ($R^2$: 0.52). Further, this effect can be considered negative, given that $C_1$ presents negative values for the vast majority of the postal codes, as pictured in Figure 4C. In addition, Figure 4C shows how the strength of the population density effect decreases (in absolute terms) as postal codes become closer to the main centers of population density. We consider the population density effect on service penetration to be concentric-positive. At the country-level, the median $C_1$ value indicates that the delay in 4G arrival decreases by a median of approximately 3 days each increase in 10 inhabitants per square kilometer. According to our data, by 2016, 4G arrived in postal codes representing 99.96 % of the population.

<u>Service quality:</u> GWR results indicate that population density can explain 31% of the observed variance ($R^2$: 0.31) in median download speeds for mobile broadband (*med_speed_mobile*). Nevertheless, this effect cannot be considered concentric since $C_1$ presents a small but significant range of negative values, as shown in Table 6. More precisely, Figure 4D depicts how the population density effect turns from positive to negative values as postal codes become closer to the main centers of population density. In fact, negative $C_1$ are only observed for a subset of postal codes with very high levels of population density. Note that this small subset is also producing the largest GWR standard residuals (Maximum Std. Res. > 3). At the country level, median $C_1$ values indicate that mobile service quality increases by a median of approximately 46.6 kbps each increase in 10 inhabitants per square kilometer. We consider the population density effect on median mobile download speed to be positive non-concentric and concave.

Summarizing, we suggest that in Finland the levels of education, mobile service penetration, spectrum usage, and mobile competition tend to increase as population density increases, thus following a positive concentric pattern. While the effect strength is weak for education and mobile competition possibly due to government intervention, it remains strong for mobile service penetration. Although the levels of mobile service quality and household income follow a concave relationship with population density, a positive concentric pattern still exists starting from a fixed distance from the city center after values are reached. In contrast, the availability of state subsidies for fiber access, private user subsidy from MNOs, and community commitment tend to decrease as population density increases, thus following a negative concentric pattern. Further, the strength of the population density effect is found to be strong for community commitment. Results for each diffusion determinant including regulation are summarized in Table 7. Note that the effect of population density is not weighted, e.g., weak, medium, strong, for those determinants that lack data.

*Table 7.Summary of population density effect on diffusion determinants*

| Diffusion determinant | Literature review | Study of Finland |
|---|---|---|
| Access network cost per user | **negative**<br>Access network cost per user<br>(Smura et al., 2007)<br>(Lyons, 2014) | **negative**<br>(no data) |
| Spectrum usage | **positive**<br>Mobile & Unlicensed<br>(Clarke, 2014)<br>(De Filippi and Tréguer, 2015) | **positive**<br>Mobile & Unlicensed<br>(Jäntti et al., 2011)<br>(Kokkoniemi and Lehtomäki, 2012) |
| Competition | **positive**<br>Operator number<br>(Grubesic, 2010)<br>(Durairajan and Barford, 2016) | **positive, weak strength***<br>*MBP_number* |
| State subsidy | **negative**<br>EU state aid regulation<br>(OECD, 2011) | **negative**<br>Broadband for all project<br>(FICORA, 2013) |
| Community commitment | **negative**<br>Social cohesion<br>(Shields, 2005) | **negative, strong strength***<br>*%_dw_ownership* |
| Education | **positive**<br>PISA secondary school<br>(OECD, 2010) | **positive, weak strength***<br>*%_bachelors* |
| Income | **concave**<br>Household income<br>(Eurostat, 2014) | **concave, medium strength***<br>*med_hh_income* |
| Private user subsidy | **negative**<br>MNO subsidy<br>(no data)<br><br>**non-concentric**<br>Mobile data sponsoring<br>(Dhanaraj Thakur, 2016) | **negative**<br>MNO subsidy<br>(Peltola and Hämmäinen, 2018) |
| Service penetration | **positive**<br>Rates for 3G and DSL<br>(EC, 2010) | **positive, strong strength***<br>*4G_diffusion_delay* |
| Service quality | **positive**<br>Download speeds<br>(OFCOM, 2017)<br>(Speedtest, 2018) | **concave, medium strength***<br>*med_speed_mobile* |

\* Effect strength (weak/medium/strong) is derived from GWR analysis

# 8. Discussion

**8.1. The role of Finnish regulation on the diffusion of mobile and fixed broadband services**

First, we suggest that the coverage obligations in 4G licenses for 800 MHz bands[4] significantly contributed to satisfying USOs since they demanded coverage of 95% and 99% of the population within 3 and 5 years, respectively. Our results show that 4G services were available in postal codes accounting for 99.97% of the population by 2016. Further, we provide evidence about the delay in 4G service arrival which increases by a median of approximately 3 days each decrease in 10 inhabitants per postal code between the auction (2013) and the obligation deadline (2016).

Second, spectrum auctions facilitated an equitable allocation of spectrum to MNOs, stimulating infrastructure-based competition (FICORA, 2017). Thus, auctions facilitated an even playing field for competitors in contrast to other EU countries where dominant operators control a majority share of the spectrum (Klemperer, 2002). Our results show that the number of mobile broadband providers is weakly affected by the population density[5].

Third, spectrum auctions enabled a low-fee allocation of spectrum in comparison to European countries, freeing financial resources for MNOs to deliver coverage obligations. Further, the NRA even allowed a joint venture between 2 out of the 3 MNOs, i.e., Telia and DNA, to comply with their 4G coverage obligations in Eastern and Northern Finland (BEREC, 2018).

Fourth, considering all previous points, we argue that the Finnish spectrum policy encouraged MNOs to satisfy USOs without the need for a USF. Spectrum auctions facilitated MNO infrastructure-based competition via equitable spectrum allocation and on-time delivery of coverage obligations via low fee spectrum licenses.

Fifth, mobile broadband was rapidly adopted as a substitute for DSL since mobile services were commercialized via flat-rate data plans[6], mobile data prices were low compared to European countries, and DSL lacked update investment (FICORA, 2016). Like in other European countries, essential facility sharing and unbundling obligations might have discouraged incumbents from upgrading DSL infrastructure (Bouckaert et al., 2010; Eskelinen et al., 2008; Pursiainen, 2007).

Sixth, mobile broadband was adopted in rural, remote areas since the education, and income levels of the population are not significantly different from those of the urban population. Our results show that the effect of population density on the number of bachelor's and the median household income have a weak and medium strength, respectively.

Seventh, although subsidies for fiber deployment existed since 2009, they did not attract investment possibly because the costs for the last 2 kilometers until the customer premises were not covered. Hence, broadband investment from nationwide operators was concentrated in mobile, given its rapid technological development and shorter pay-back time, as observed in other countries (Briglauer, 2014).

Eight, we argue that since nationwide operators did not sufficiently invest in fiber access, entrepreneurial action from consumers was triggered, leading to the emergence of fiber consumer cooperatives. Consumer self-organization could be caused by a rising demand for bandwidth which was not satisfied by mobile services nor by the minimum download speeds of USOs. This emergence can be compared to those of European wireless community networks. We provide a plausible explanation for this emergence by showing that when population density decreases, the levels of mobile service quality decrease and community commitment increase.

---

[4] 800 MHz frequencies can provide wider coverage at a lower cost than other higher frequencies commonly used for broadband provision.

[5] Since in Finland DSL-based fixed access was provided by regional operators, the deployment of mobile broadband by these operators did not cannibalize the existing offering when deployed beyond their regional coverage.

[6] Flat-rate data pricing was first introduced by a single operator, thus forcing competitors to adopt it as well. This operator had the smallest share of the Finnish market and did not operate in any other European market.

### 8.2. Regulation for 5G

Our statistical results show two trends when population density decreases. On the one hand, levels of service penetration, service quality, and competition tend to decrease. On the other hand, levels of unused spectrum and community commitment tend to increase. Since community commitment is a relevant prerequisite for community operator emergence, we suggest that there is a market opportunity for 5G local operators in low populated areas.

Based on learnings from the Finnish regulation and considering the high costs of 5G deployment, we suggest regulators implementing market-driven policies for 5G to stimulate demand-driven investment in commercially underserved areas by taking advantage of community commitment. For example, community operators could play a role in the deployment of shared, small cell networks, which can accelerate 5G coverage into rural areas as shown by recent studies (Oughton and Frias, 2017). In our view, stimuli should include the allocation of local spectrum licenses, given the lower access cost per user of mobile services. When a new broadband service is launched, nationwide rules could account for local actors (including consumers) and their aggregated investment capacity. Thus, service diffusion may start from centers of population with a lower population density than in 4G, thus reducing the diffusion delay in less populated areas.

However, in Finland the 3.5 GHz band has been fully allocated to MNOs, preventing an increase in local competition. Nevertheless, according to the spectrum license agreement, MNOs have an obligation to offer a local license if they cannot serve a service request from a local customer.

## 9. Conclusions

This article analyzes the role of the Finnish regulation in achieving the broadband penetration goals defined by the NRA. We propose the concentric patterns of broadband diffusion, defining a spatial relationship between population density and the main determinants of broadband diffusion. We measure these patterns through a GWR analysis including postal code data from mainland Finland during 2013-2016. First, we show that spectrum auctions with coverage obligations, including the 800Mhz auction, accelerated mobile service diffusion since 4G arrived in postal codes representing 99.96 % of the population by 2016. Second, we find indications that the design of 4G spectrum auctions, which facilitated the allocation of an equal amount of spectrum to competing operators, achieved high levels of mobile service competition throughout Finland. Third, we suggest that spectrum auctions facilitated the allocation of spectrum at relatively low fees, encouraging MNOs to employ financial resources to comply with coverage obligations. Based on these three points, we argue that spectrum policy supported the broadband penetration goals of Finland, encouraging MNOs to satisfy USOs without the need for a USF. Finally, given that fiber consumer cooperatives emerged in rural areas, we indicate that state subsidies stimulated demand-driven investment. We provide a plausible explanation for this emergence by showing that levels of unused spectrum and community commitment increase with decreasing population density. Based on these findings, we recommend NRAs adopting market-driven policies for 5G to account for local actors, including consumers and their aggregated investment capacity. Thus, service diffusion can start from centers of population with a lower population density than in 4G, thus reducing the diffusion delay in less populated areas. Finally, we recommend three regulatory actions that predominantly apply to OECD countries with market-driven regulatory policies, strong concentric patterns (e.g., caused by population dispersion, weak regulatory mitigation), and moderate levels of income and education among the rural population. Next, we list the above-mentioned three regulatory actions:

1. <u>Local competition should be promoted with local 5G spectrum licenses relying on local community commitment</u>. Frequency bands such as 3.5 GHz and higher should be allocated locally or regionally to promote local competition in mobile and fixed-wireless services (e.g., through the entry of new 5G micro-operators, indoor and outdoor). This regional allocation is likely to promote spectrum utilization and hinder underutilization as observed, e.g., in the case of the 2.6 GHz band in Finland. The recent allocation of the 3.5 GHz band to Finnish MNOs might prevent the above-mentioned increase in local competition.

2. <u>Local fiber access diffusion should be promoted relying on local community commitment</u>. The challenging problem of fiber access monopoly can be alleviated with a mix of access-based competition, wholesale-only requirement, and price caps.

3. <u>Investment in commercially underserved areas should be incentivized via subsidies and long-term loans relying on community commitment.</u> The early deployment of new technology should be incentivized to accelerate rural penetration, increase competition, and improve the nationwide social surplus via the positive network effects of late adopters.

**10. Limitations and future work**

This study is limited by the nature of the Paavo and Netradar datasets. Regarding the Paavo dataset, the GWR may be affected by the absence of socio-economic data for less than 150 postal codes, which have very low population density due to privacy reasons. GWR may also be affected by the changing size of postal codes, i.e., larger areas for lesser populated postal codes and smaller areas for higher populated postal codes, as commonly found in census datasets. As a result, the existence of significant population density centers which are located in large postal codes might be hidden. Hence, values of diffusion determinants that belong to these centers (e.g., high download speeds from a base station in the city center) are instead associated with averaged densities. Future work may also utilize population density data with 1 square kilometer resolution. Regarding the Netradar dataset, the GWR may be affected by the lack of data from less populated postal codes since in these postal codes there potentially are less Netradar users, and they have a larger area to cover. As a result, contribution from some of these postal codes might not be included in GWR coefficients, which may slightly modify minimum or maximum values. Nevertheless, the Netradar dataset included a substantial number of measurements from more than 2,000 postal codes already in 04/2013, which is the starting date for the LTE diffusion study. Exact number of used measurements is 972,890 for *MBP_number* and *med_speed_mobile*, and 850,175 for *4G_diffusion_delay*. In addition, the GWR may also be affected by sample selection effects since Netradar users voluntarily install the application without any compensation.

This study is also limited by the employed method. The study measures the population density effect through linear models. Although the explanatory power of population density could be better captured via curvilinear models, the goodness-of-fit comparison between GWR models is not possible since the required logarithmic transformations imply different error model assumptions.

This study has estimated the sole contribution of population density to the variance of diffusion determinants, thus limiting the explanatory power of GWR models. New models could study individual regulatory policies, albeit it might be challenging to find proxy variables at the postal code level. Future work may reproduce the study for other countries, enabling the comparison of regulatory mitigation.

## Acknowledgments

The authors would like to thank Alexandr Vesselkov and Matti Kilkki for their valuable comments and discussions. We also thank Benjamin Finley for his support during the analysis of Netradar data. The work was supported by the EC H2020 RIFE project Grant No. 644663. Andrés Arcia-Moret has been funded by the projects Network as a Service (EP/K031724/2) and SPARE (EP/R511675/1 NRAG/527).